\documentclass[%
 aip,
 amsmath,amssymb,
 reprint,%
]{revtex4-1}

\usepackage{graphicx}
\usepackage{dcolumn}
\usepackage{bm}

\usepackage[utf8]{inputenc}
\usepackage[T1]{fontenc}
\usepackage{mathptmx}
\usepackage{etoolbox}

\usepackage{caption}
\usepackage{subcaption}
\usepackage{adjustbox} 
\usepackage{multirow} 

\usepackage{array}
\renewcommand{\arraystretch}{1.6}
\newcommand{\Pe} {\text{Pe}}

\newcommand{\abs}[1] {\lvert #1 \rvert}

\makeatletter
\def\@email#1#2{%
 \endgroup
 \patchcmd{\titleblock@produce}
  {\frontmatter@RRAPformat}
  {\frontmatter@RRAPformat{\produce@RRAP{*#1\href{mailto:#2}{#2}}}\frontmatter@RRAPformat}
  {}{}
}%
\makeatother
\begin{document}

\preprint{AIP/123-QED}

\title[Radial Variations in Residence Time Distribution for Pipe Flows]{Radial Variations in Residence Time Distribution for Pipe Flows}
\author{Etienne Boulais}
\author{Richard D. Braatz}%
    \email{braatz@mit.edu}
\affiliation{ 
Department of Chemical Engineering, MIT
}%

\date{\today}

\begin{abstract}
Suspensions of low-diffusing particles in pipe flows exhibit a difference in age at different radial positions. Particles near the channel walls have higher residence times than the cross-sectional average. We quantify this effect using Monte-Carlo simulations, and show the existence of two different regimes: a "transitional" regime where delay compounds with channel length, and a "far-field" regime where diffusion counterbalances advection. The results presented therein can be used to quantify residence time distributions near the walls of the tube. This effect is important to consider in experiments involving the kinetics of nanometer-scale particles using modern inline analytical tools. This work also provide a radially resolved extension of classical Taylor dispersion results.
\end{abstract}

\maketitle




\section{Introduction}


It is well-known that, in channel flows at high Peclet number, the interaction of lateral diffusion with the parabolic flow profile causes a broadening of concentration clouds. The problem was initially studied by Taylor \cite{taylor1953dispersion}, who showed that the far-field cross-sectionally averaged concentration cloud tends to a Gaussian profile with an effective diffusivity scaled with $\Pe^2$; then by Aris \cite{aris1956dispersion}, who corrected Taylor's initial estimate for the effective diffusivity.
%
Following this pioneering work, a number of different authors studied the problem.
Applications of the model to the measurement of average velocities in pipes has been discussed by Levenspiel \cite{levenspiel1957notes}.
In a series of papers, Gill and Ananthakrishnan \cite{ananthakrishnan1965laminar} \cite{gill1971dispersion} extended the results to cases of different inlet conditions and velocity profiles.
Chatwin\cite{chatwin1970approach} studied the transient behavior of the concentration cloud in the near-field and its convergence to Taylor's Gaussian profile.
In the mid-2000s, some interest in the problem was raised again in the context of microfluidic systems \cite{beard2001taylor} \cite{ajdari2006hydrodynamic}, and more recently, experimental work is connecting Taylor dispersion with problems transport in biological systems \cite{vilquin2021time}. 

Whereas most work is concerned with the temporal evolution of the concentration of a cloud of tracer, in other words with the distribution of positions of individual particles at fixed times, the problem of residence time distribution asks instead how particles' ages are distributed at a set position in the system.
Residence time distribution measurements are common in chemical engineering applications \cite{levenspiel1998chemical}, where techniques for their determination are well-established, and easily applicable to almost any continuous systems. Problems of residence time distributions arise naturally in continuous systems, where measurement equipment is distributed at set positions along a process pipeline, sampling particles of different age at a set position rather than sampling every position in the system at a set instant.
The interaction of Taylor dispersion with residence time distributions has been much less studied than the concentration problem. Houseworth \cite{houseworth1984shear} gave some analytical scalings, as well as numerical results based on Monte-Carlo simulations for the near-field problem. However, the work only considered cross-sectionally averaged quantities, and did not report variations of residence times in the radial direction.

Such variations of residence time in the radial direction become important when studying complex kinetics involving colloids or nanoparticles evolving in pipe flow. For example when probing the self-assembly of lipid nanoparticles \cite{udepurkar2025structure}, polymeric particles \cite{hayward2010tailored}, or other small lipid vesicles \cite{leng2003kinetics}. In such a system, because of Taylor dispersion, the relative age of particles at different radial positions in the tube will be quite different. If the studied systems are far from equilibrium, these differences in residence time may translate to significant differences in particle properties (radius, morphology, charge, etc.), depending on the kinetics under study.

In this paper, we show how residence times vary along the cross-section in both 2D semi-infinite channels and cylindrical tubes. We show that particles near the wall have higher median residence times, and quantify the effect using Monte-Carlo simulations.
This effect becomes important to consider when using noninvasive analytical technologies to characterize particles in pipe flows, in cases where the system's penetration is finite and doesn't encompass the whole channel. For example, flow dynamic light scattering (DLS) systems \cite{chowdhury1984application} \cite{leung2006particle} \cite{besseling2019new} are used to characterize nanoparticles' size distribution by interrogating a small area near the edge of a channel.
Other inline technologies include microscopy systems with finite penetration depth near the tube's surface, optical coherence tomography systems \cite{weiss2013localized}, or spectroscopic systems \cite{bakeev2010process}, to name a few.
In such situations, the particles interrogated are not necessarily representative of the whole cross-section but instead have a certain delay over the rest, which we quantify in this article. We show that the accumulated delay can be easily be of the order of seconds to minutes, depending on the experiment and solute-solvent properties. This effect can be especially significant when studying systems that have kinetics comparable to that delay, for example, in applications related to industrial crystallization or nanoparticle growth.





\section{Theory}


\begin{figure}
	\centering
    \includegraphics[width=0.4\textwidth]{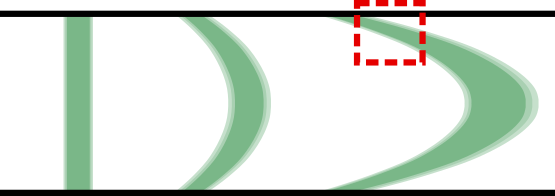}

	\caption{Schematic of the temporal evolution of a thin plug under the conditions of Taylor dispersion}
	\label{fig_taylor_cartoon}
\end{figure}

The probability density $P$ for a particle's position is governed by the advection-diffusion equation, 
\begin{equation}\label{eq_advection_diffusion}
    \frac{\partial P}{\partial t} = D \nabla^2 P - \vec{u} \cdot \nabla P,
\end{equation}
where $D$ is the diffusion coefficient and $\vec{u}$ is the velocity field.
For flow in a cylindrical channel of circular cross-section, the velocity follows the parabolic flow profile $\vec{u} \left( z \right) = 2 V \!\left( 1 - \frac{r^2}{a^2} \right)$, where $a$ is the channel's radius and $V$ is the flow's average velocity.
Equation \ref{eq_advection_diffusion} can be nondimensionalized using the scalings 
\begin{equation}\label{eq_scalings}
    \bar{z} = \frac{D z}{V a^2}, \quad \bar{t} = \frac{D t}{a^2}, \quad \eta = \frac{r}{a}.
\end{equation}

In cylindrical coordinates, the equation becomes
\begin{equation}
    \frac{\partial P}{\partial \bar{t}} + 2 ( 1 - \eta^2 ) = \frac{1}{\Pe^2} \frac{\partial^2 P}{\partial z^2} + \frac{\partial^2 P}{\partial \eta^2} + \frac{1}{\eta} \frac{\partial P}{\partial \eta} + \frac{1}{\eta^2}\frac{\partial^2 P}{\partial \theta^2}
\end{equation}
where $\Pe = \frac{V a}{D}$ is the Peclet number. At high values of the Peclet number, the retrodiffusion term $\frac{1}{\Pe^2} \frac{\partial^2 P}{\partial z^2}$ can be neglected. In addition, if only the variation in the radial distance is of interest, the equation can be averaged over the angular coordinate $\theta$ to yield
\begin{equation}\label{eq_advection_diffusion_avg}
    \frac{\partial P^*}{\partial \bar{t}} + 2 ( 1 - \eta^2) \frac{\partial P^*}{\partial \hat{z}} = \frac{\partial^2 P^*}{\partial \eta^2} + \frac{1}{\eta} \frac{\partial P^*}{\partial \eta}
\end{equation}
where $P^* = \frac{1}{2 \pi}\int_0^{2 \pi} P d\theta$.
%
%
$\Pe$ is absent from equation \ref{eq_advection_diffusion_avg}, which means that, in the high Peclet regime, the problem only has to be analyzed once and the results can then be transferred to any geometry through the scalings in equation \ref{eq_scalings}.

\newcommand*{\citen}[1]{%
  \begingroup
    \romannumeral-`\x 
    \setcitestyle{numbers}%
    \cite{#1}%
  \endgroup   
}

\section{Monte-Carlo Simulations}

In order to study the interplay of residence time distribution and radial position of the particles, we implement the Monte-Carlo method described in Ref.\  \citen{houseworth1984shear}. The trajectory of individual particles are simulated in a series of discretized steps.
Each timestep is subdivided into an advection step, where the axial position of the particle is updated,
\begin{equation}
    \hat{z}_{i+1} = \hat{z}_{i} + u ( \eta ) \Delta \hat{t},
\end{equation}
and a diffusion step, where the particle is allowed to move along the tube's cross-section,
\begin{equation}
    \eta_{i+1} = f ( \eta_i, \Delta \hat{t} ),
\end{equation}
where the new radial position $\eta_{i+1}$ has to be chosen using an appropriate probability distribution. If small enough timesteps are used, then the diffusion step is approximately decoupled from the advection step. In that case, the probability distribution $\gamma \left( \eta_{i+1}, \Delta t; \eta_i \right)$ for the new radial position of the particle $\eta_{i+1}$, given the current position $\eta_i$, is given by the impulse response of the diffusion equation, 
\begin{equation}
    \frac{\partial \gamma}{\partial t} = \frac{\partial^2 \gamma}{\partial \eta} + \frac{1}{\eta} \frac{\partial \gamma}{\partial \eta},
\end{equation}
with boundary condition $\frac{\partial \gamma}{\partial \eta} = 0$ at $\eta = 1$ and initial condition
\begin{equation}
    \gamma \left( \eta_{i+1}, 0; \eta_i \right) = \frac{1}{\eta_i} \delta ( \eta_{i+1} - \eta_i ),
\end{equation}
which corresponds to an infinitely thin ring source located at radial position $\eta_i$, where $\delta ( x )$ is the Dirac delta function. The solution to this problem is well-known and can be found in handbooks on partial differential equations \cite{polyanin2001handbook} or in classic treatises on diffusion \cite{crank1979mathematics}. For a given timestep $\Delta t$, we find that
\begin{eqnarray}
    \gamma \left( \eta_{i+1}, \Delta t; \eta_i \right) = \qquad \qquad\qquad\qquad\qquad\qquad\qquad\nonumber\\ 2\! \left(\! 1 + \sum_{N=1}^{\infty} \exp \left( - \alpha_N^2 \Delta \hat{t}^2 \right) \frac{J_0 ( \eta_i \alpha_N ) \, J_0 ( \eta_{i+1} \alpha_N )}{J_0^2 ( \alpha_N )} \right)\!,
\end{eqnarray}
which can be integrated to yield the cumulative probability function
\begin{eqnarray}
    S \left( \eta_{i+1}, \Delta t; \eta_{i} \right) = \eta_{i+1}^2\qquad \qquad \qquad\qquad\qquad\qquad\nonumber\\
+ 2 \eta_{i+1}\! \sum_{N=1}^{\infty} \! \left( \!\frac{\exp( - \alpha_N^2 \Delta \hat{t} )}{\alpha_N} \frac{J_1 ( \eta_{i+1} \alpha_N) J_0( \eta_i \alpha_N )}{J_0^2 (\alpha_N)} \!\right),
\end{eqnarray}
where $J_0$ and $J_1$ are Bessel functions of the first kind \cite{abramowitz1968handbook}, and the constants $\alpha_N$ are the positive zeros of the Bessel function $J_0$.
The function $S$ can be numerically inverted to obtain
\begin{equation}
    \eta_{i} = g ( S, \eta_{i-1}, \Delta \hat{t} ).
\end{equation}

As a cumulative probability function, $S$ is uniformly distributed between 0 and 1. Given a random number generated from a uniform distribution between 0 and 1, $g( S, \eta_{i-1}, \Delta \hat{t})$ yields a value for the next radial position, following the correct statistical distribution. To avoid having to invert $S$ at every step, which would be unnecessarily slow, we precompute an interpolant for $g$ on a fine grid covering all possible values of $S$ and $\eta_{i-1}$.

We simulate two different initial conditions. The first one is an infinitely thin area source (or plug) with uniform distribution on the cross-section at $\hat{z} = 0$, which is done by taking
\begin{equation}
    \eta_{0} = \sqrt{S}
\end{equation}
where $S$ is a random number uniformly distributed between 0 and 1. The square root accounts for the radial distortion of the area element in cylindrical coordinates. 
We also simulate a uniform flux of particles along the cross-section (where more particles appear near the middle of the channel due to the higher flow rate there. Given a random number $S$, the initial radial position of a particle is then \cite{houseworth1984shear}
\begin{equation}
    \eta_0 = \sqrt{1 - \sqrt{S} }.
\end{equation}

The constant flux condition is more representative of a situation where a well-mixed solution is fed into the pipe than the somewhat artificial constant concentration condition. In both cases, the results are qualitatively similar. We present all graphs for both initial conditions in Appendix \ref{appendix_additional}. 

\subsection{2D Channel}

In addition to cylindrical channels, we also analyze the problem for unidirectional flow in a semi-infinite gap between two plates. This problem can come in handy when studying microfluidic channels and chambers which, because of the way they are fabricated, are often correctly modeled as semi-2D systems \cite{boulais20232d}. The procedure for analyzing the semi-infinite system is exactly the same as for a cylindrical one. In that case, the nondimensional advection-diffusion equation (neglecting longitudinal diffusion) takes the form
\begin{equation}
    \frac{\partial P^*}{\partial \hat{t}} + \frac{3}{4} ( 1 - \hat{y}^2 ) \frac{\partial P^*}{\partial \hat{z}} = \frac{\partial^2 P^*}{\partial \hat{y}^2}
\end{equation}
where $\hat{y}$ is the dimensionless vertical position, with walls situated at $\hat{y} = \pm 1$.
The diffusive step in the Monte-Carlo algorithm is done the same way as before, this time using the simpler form of the diffusion equation,
\begin{equation}
    \frac{\partial \gamma}{\partial t} = \frac{\partial^2 \gamma}{\partial \hat{y}^2},
\end{equation}
with Green's function \cite{polyanin2001handbook}
\begin{eqnarray}
    \gamma \left( \hat{y}_{i+1}, \Delta t; \hat{y}_i \right) = 1\qquad\qquad\qquad\qquad\qquad\qquad\qquad \nonumber\\+ 2 \sum_{N=1}^{\infty} \cos ( N \pi \hat{y}_{i+1} ) \cos ( N \pi \hat{y}_i) \exp \!\left( -N^2 \pi^2 \Delta t\right)
\end{eqnarray}
and cummulative probability distribution
\begin{eqnarray}
    S \left( \hat{y}_{i+1}, \Delta t ; \hat{y}_i \right) = \hat{y}_i \qquad\qquad\qquad\qquad\qquad\qquad\qquad\nonumber\\+ \frac{2}{\pi} \sum_{N=1}^{\infty} \frac{1}{N} \sin ( N \pi \hat{y}_{i+1} ) \cos ( N \pi \hat{y}_i) \exp\! \left( -N^2 \pi^2 \Delta t\right).
\end{eqnarray}

Initial conditions in the case of the 2D channel geometry are much simpler. For a uniform area source the initial vertical position $\hat{y}_0$ is simply obtained by generating a uniform random number. 
For uniform flux, we use the cummulative probability distribution (for the top half-channel)
\begin{align}
    S ( \hat{y}_0 )&= \frac{3}{2} \int_0^{\hat{y}_0} \left( 1 - \xi^2 \right) d \xi, \\
   S & = \frac{3}{2}\! \left( \hat{y}_0 - \frac{1}{3} \hat{y}^3_0 \right).
\end{align}
Solving for $\hat{y}_0$ and generating a uniform number between 0 and 1 for $S$ gives the proper constant flux distribution in the top half channel.

\section{Results}

\subsection{2D Channel Geometry}

\begin{figure*}
	\centering
    \begin{subfigure}{0.32\textwidth}
		\includegraphics[width=\textwidth]{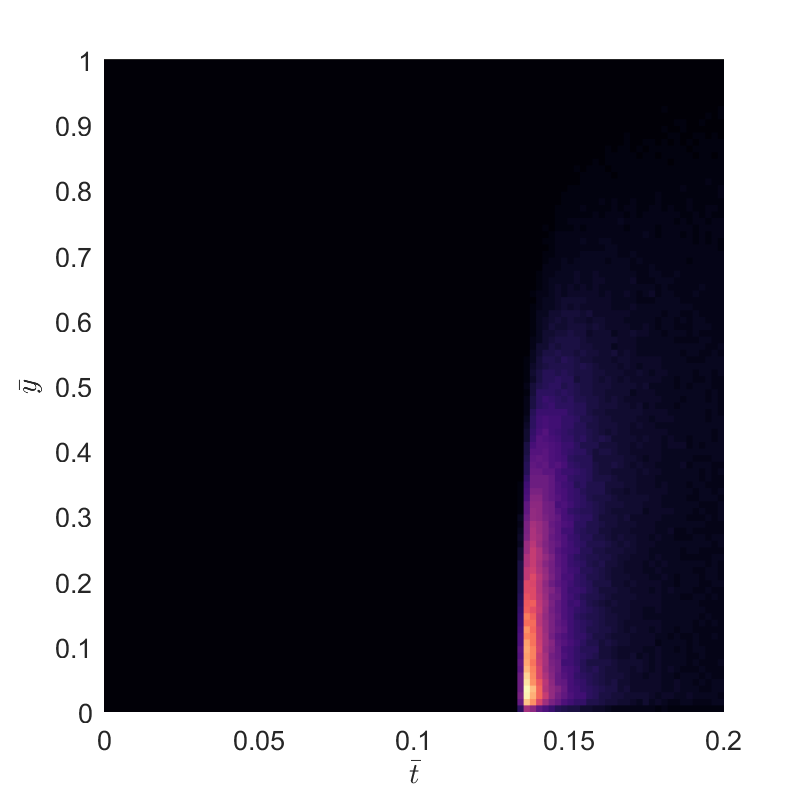}
        
        \vspace{-0.2cm}
        
		\caption{$\hat{z} = 0.1$}
		\label{subfig_histogram_02}
	\end{subfigure}
	\begin{subfigure}{0.32\textwidth}
		\includegraphics[width=\textwidth]{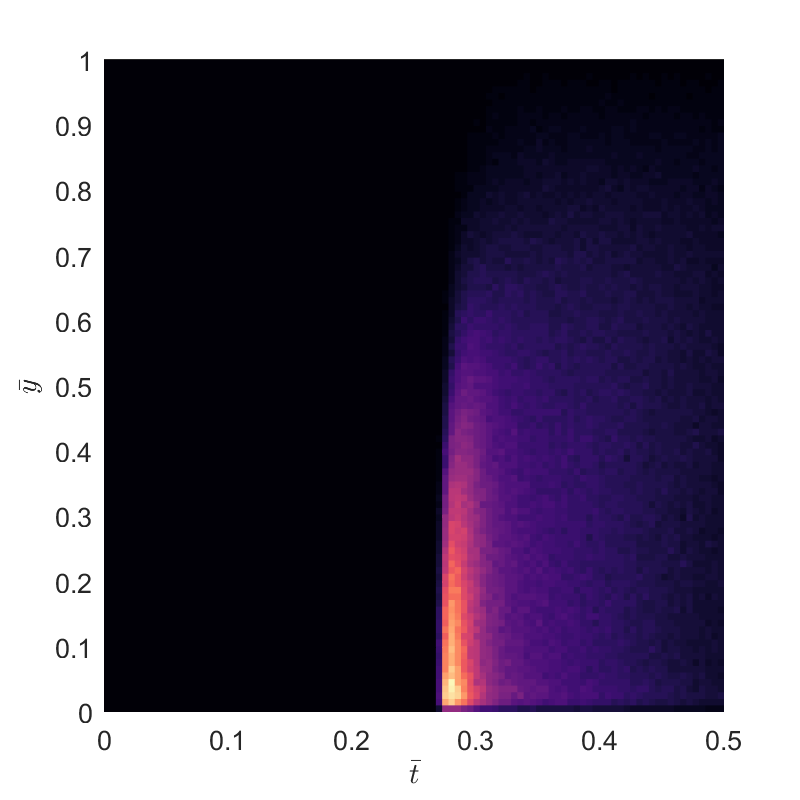}
        
        \vspace{-0.2cm}
        
		\caption{$\hat{z} = 0.2$}
		\label{subfig_histogram_04}
	\end{subfigure}
    \begin{subfigure}{0.32\textwidth}
		\includegraphics[width=\textwidth]{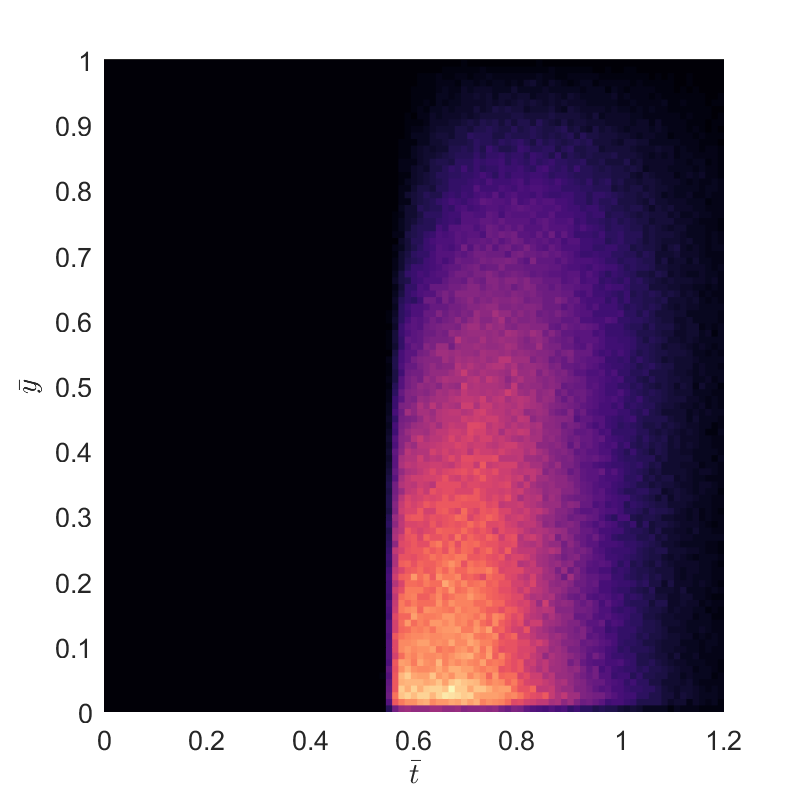}

        \vspace{-0.2cm}
        
		\caption{$\hat{z} = 0.4$}
		\label{subfig_histogram_08}
	\end{subfigure}

\vspace{-0.2cm}

	\caption{2D histogram of vertical position $\hat{y}$ and age $\hat{t}$ of particles crossing different channel lengths $\hat{z}$. Constant flux source.}
	\label{fig_histogram}
\end{figure*}

\begin{figure}
	\centering
    \includegraphics[width=0.4\textwidth]{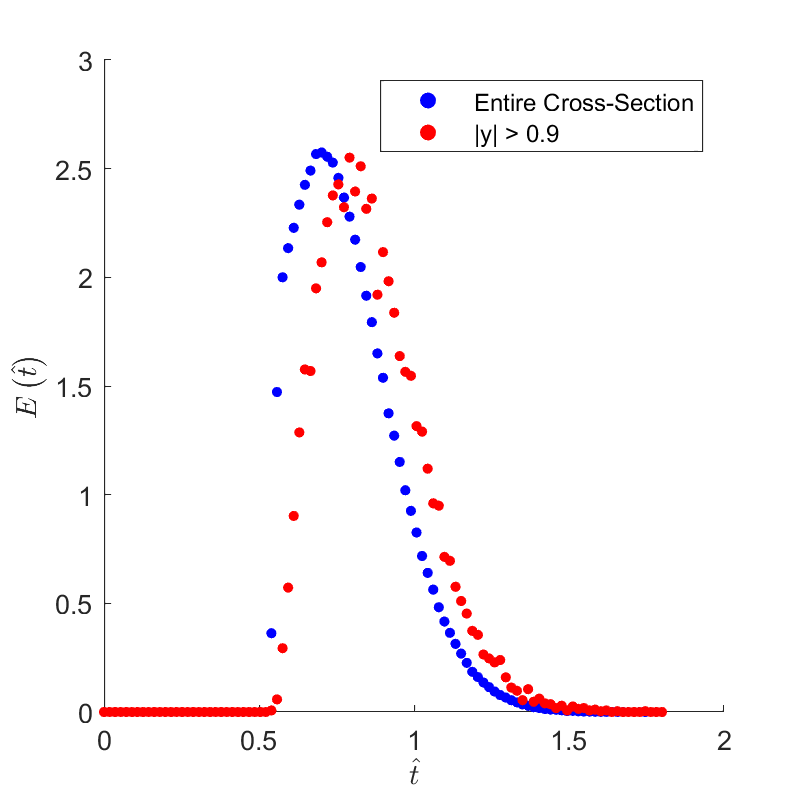}

\vspace{-0.3cm}

	\caption{Residence time distribution $E ( \hat{t} )$ curve at $\hat{z} = 0.4$ for the entire channel cross-section (blue), as well as equivalent curve counting only particles for which $\abs{\hat{y}} > 0.9$ (red) The red curve amounts for a smaller number of particles, but for comparison both curves are normalized to have unit area.}
	\label{fig_single_rtd}
\end{figure}

\begin{figure*}
	\centering
    \begin{subfigure}{0.45\textwidth}
		\includegraphics[width=\textwidth]{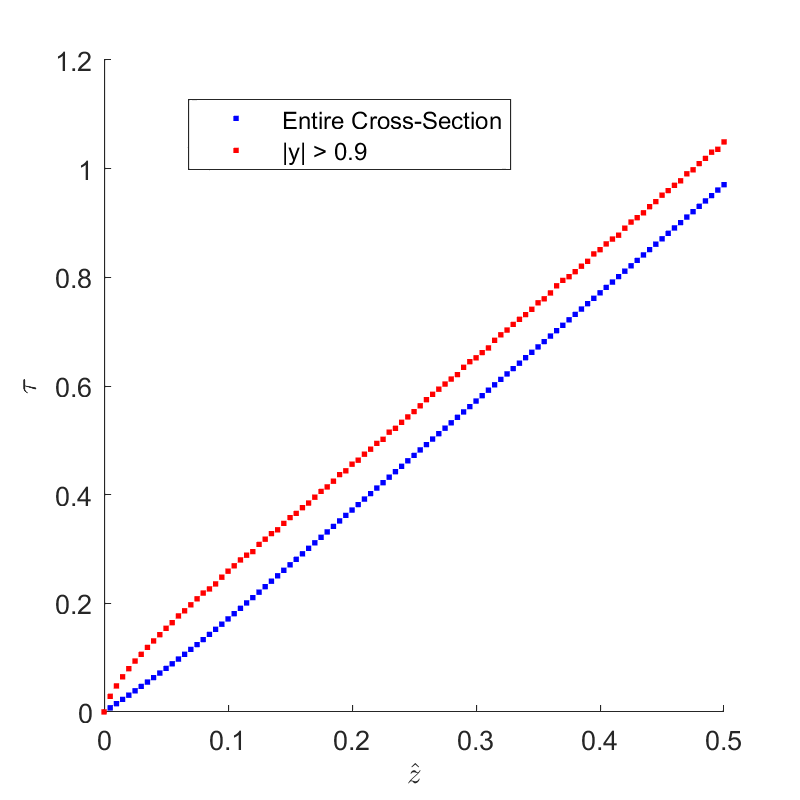}

        \vspace{-0.2cm}
        
		\caption{}
		\label{subfig_rtd_avg}
	\end{subfigure}
	\begin{subfigure}{0.45\textwidth}
		\includegraphics[width=\textwidth]{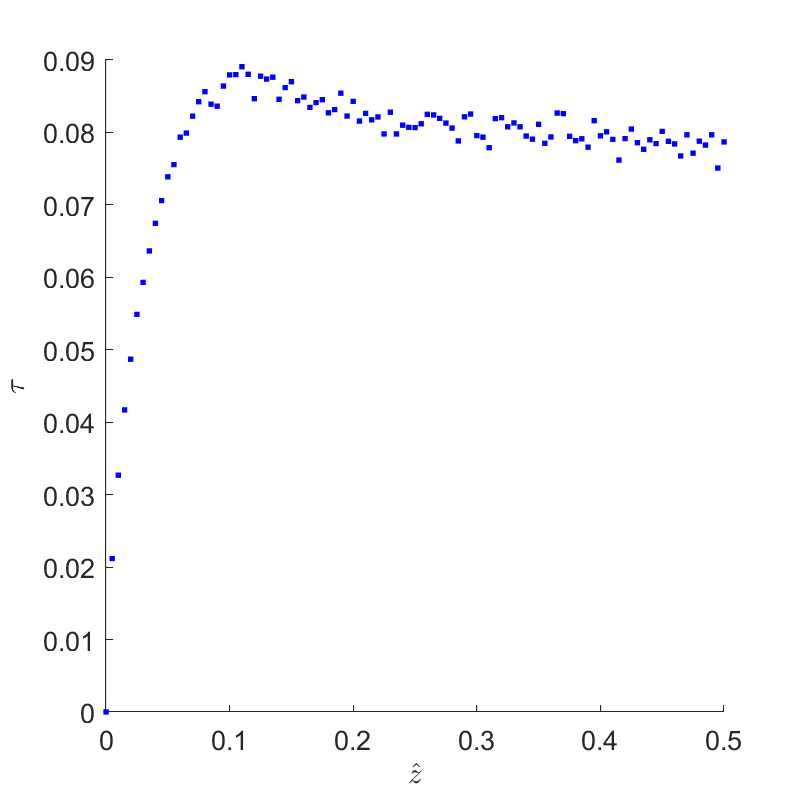}
        
        \vspace{-0.2cm}
        
		\caption{}
		\label{subfig_rtd_average_diff}
	\end{subfigure}

        \vspace{-0.2cm}
        \caption{(\subref{subfig_rtd_avg}) Median residence time for particles at different axial distances down the channel for entire cross-section (blue), and for particles with $\abs{\hat{y}} > 0.9$ (red). (\subref{subfig_rtd_average_diff}) Difference in median residence time between both curves.}
	\label{fig_rtd_comparison}
\end{figure*}

\begin{figure*}
	\centering
    \begin{subfigure}{0.45\textwidth}
		\includegraphics[width=\textwidth]{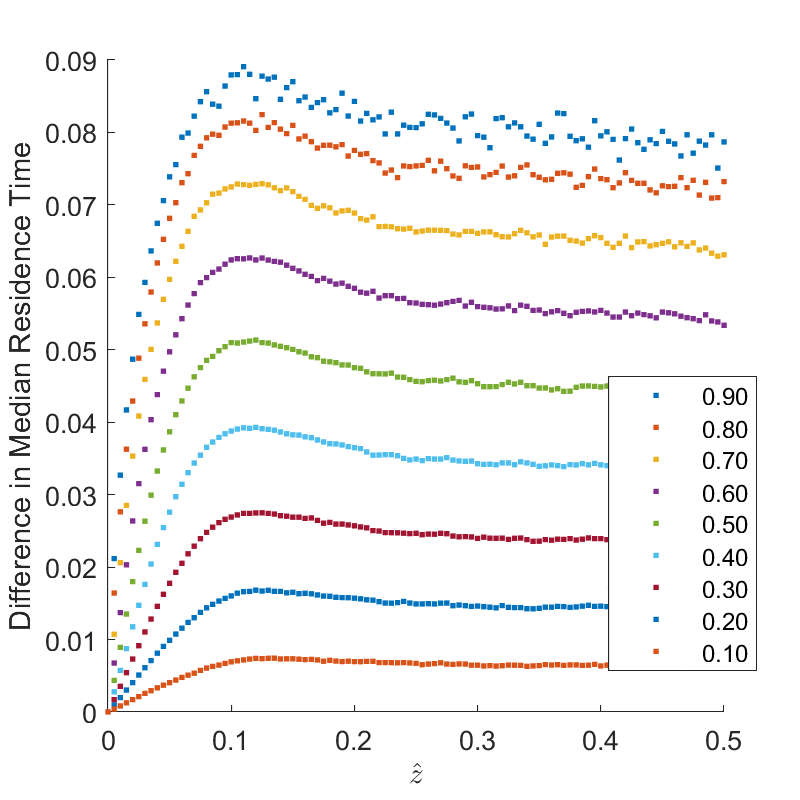}
        
        \vspace{-0.2cm}
        
		\caption{}
		\label{subfig_rtd_depths}
	\end{subfigure}
	\begin{subfigure}{0.45\textwidth}
		\includegraphics[width=\textwidth]{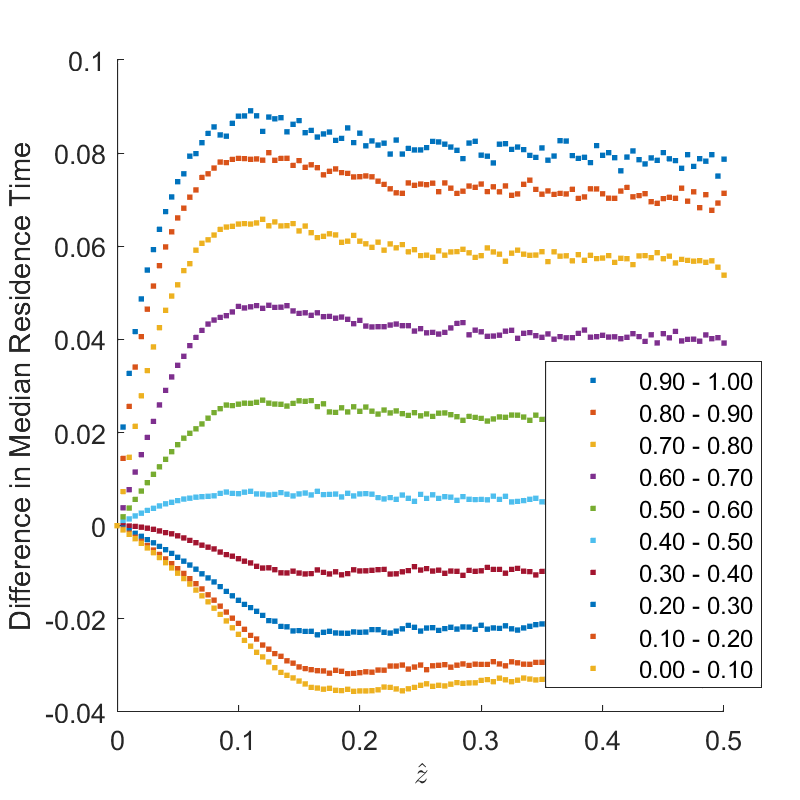}
        
        \vspace{-0.2cm}
        
		\caption{}
		\label{subfig_rtd_slices}
	\end{subfigure}

        \vspace{-0.2cm}
        
	\caption{Difference in dimensionless median time for different penetration depths $\abs{\hat{y}} < k$ (a), as well as for different slices $k_1 \leq \abs{\hat{y}} < k_2$ (b). Constant flux source.}
	\label{fig_rtd_depths}
\end{figure*}

We begin by analyzing the 2D channel geometry, which is somewhat simpler than the cylindrical channel. Results for the cylindrical case are presented in Section \ref{sec_results_cylinder}.
We simulate the trajectory for 1 million particles with a time step of $\Delta \hat{t} = 0.001$ and record both their vertical position $\hat{y}$ and their age $\hat{t}$ as they cross set axial distances $\hat{z}$ in the channel. For each axial position, the distribution of $\hat{y}$ and $\hat{t}$ can then be plotted in a 2D histogram, examples of which are given in Fig.\  \ref{fig_histogram}.
%
%
This 2D histogram contains both information on the radial concentration profile (the sum along the $\hat{t}$ direction is proportional to the time-integrated concentration profile) and the residence time distribution (obtainable by summing along the $\hat{y}$ direction) for different axial distances in the tube.
However, this type of plot also contains additional information on the interplay of radial position, concentration, and residence times, seen through the correlations of the $\hat{t}$ and $\hat{y}$ distributions.
In order to visualize some of this interplay, we generate an equivalent of the residence time distribution $E ( \hat{t})$ curve that only counts particles whose radial position $\hat{y}$ is above a certain threshold (corresponding to the particles that are a given distance from the walls of the channel). This is shown in Fig.\  \ref{fig_single_rtd} for a penetration depth of $0.1$. We can see that the age of the particles near the wall is, unsurprisingly, slightly higher than those over the entire cross-section, which is to be expected due to the lower velocities near the wall.
To compare how this lag evolves with increasing channel length, we compute the median particle age for each distribution, corresponding to the time $\tau$ for which 
\begin{equation}
    \int_0^{\tau} E ( \hat{t} ) d \hat{t} = \frac{1}{2}.
\end{equation}
%
This median time is plotted in Fig.\ \ref{subfig_rtd_avg} for both the entire channel cross-section, and for the particles for which $\abs{\hat{y}} > 0.9$. We observe that the median age of the particles near the wall initially increases relative to the whole channel, but then stabilizes and grows at the same pace. This effect is illustrated in Fig.\ \ref{subfig_rtd_average_diff} which plots the difference between the median age of particles near the wall and those in the entire channel. The resulting curve exhibits two distinct regimes, one for low values of $\hat{z}$ where the delay due to the lower average velocity near the wall compounds more or less linearly, and a second one at high $\hat{z}$ where a maximum lag is reached, and the difference in median age between the two distributions remains constant. This region can be interpreted as the one where an equilibrium is reached between the delay caused by lower average velocities near the walls and diffusive motion exchanging particles between the wall region and the middle of the channel.
%
%
As an additional feature of Fig.\  \ref{subfig_rtd_average_diff}, we see that the difference in delay initially "overshoots" the equilibrium value, before slowly settling back down at higher values of $\hat{z}$. This is due to the fact that parts of the channel that are near the walls settle to their equilibrium configuration faster than parts near the middle (an effect also visible in Fig.\  \ref{subfig_rtd_slices}, discussed below), thus initially biasing the delay towards a value higher than the final value.
We next investigate the variation of this delay in the radial direction, which is illustrated in Fig.\ \ref{fig_rtd_depths}. Figure \ref{subfig_rtd_depths} shows the difference in median residence time between the whole channel and particles for which $\hat{y} > k$ for different values of $k$ ($k = 0$ corresponding to the entire distance between the channel's center and the wall). In each case, the qualitative behavior is the same as observed previously, with lag compounding approximately linearly, then reaching a transition region, then settling on a constant value at higher values of $\hat{z}$.
In subfig \ref{subfig_rtd_slices}, we compare the difference in median residence time between the whole channel and a thin slice $k_0 \leq \hat{y} < k_1$. We can see that, on average, particles for which $\abs{\hat{y}} > 0.5$ are lagging behind the cross-sectional average, while particles for which $\abs{\hat{y}} < 0.4$ have a smaller median age than the cross-sectional average. The switch between positive and negative lag happens between the $0.3 \leq \hat{y} < 0.4$ and $0.4 \leq \hat{y} < 0.5$ curves.

\subsection{Cylindrical Channel}\label{sec_results_cylinder}



\begin{figure*}
	\centering
    \begin{subfigure}{0.32\textwidth}
		\includegraphics[width=\textwidth]{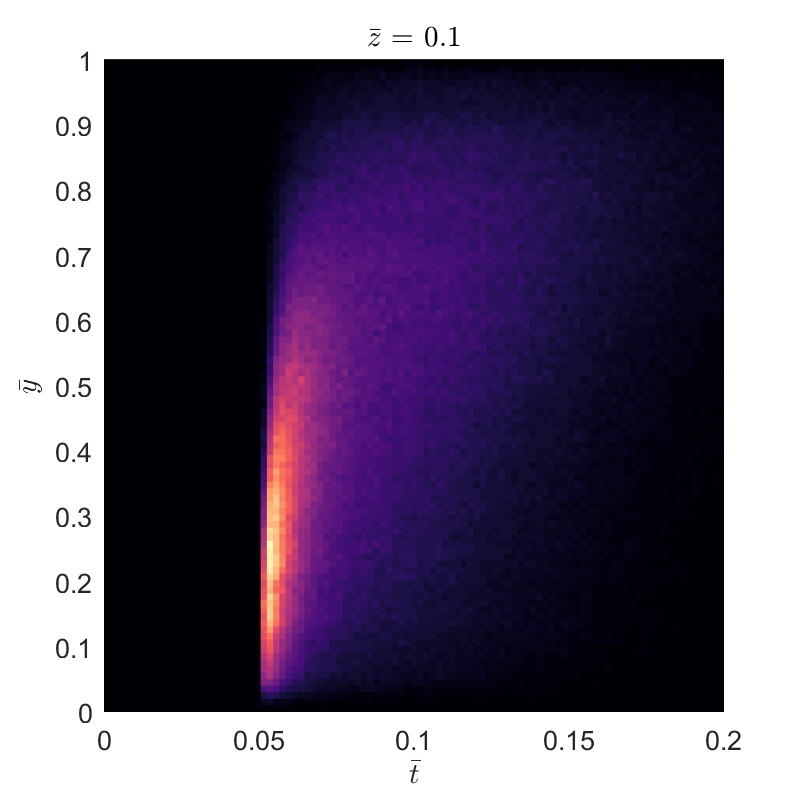}

        \vspace{-0.2cm}
        
		\caption{$\hat{z} = 0.1$}
		\label{subfig_histogram_cylinder_01}
	\end{subfigure}
	\begin{subfigure}{0.32\textwidth}
		\includegraphics[width=\textwidth]{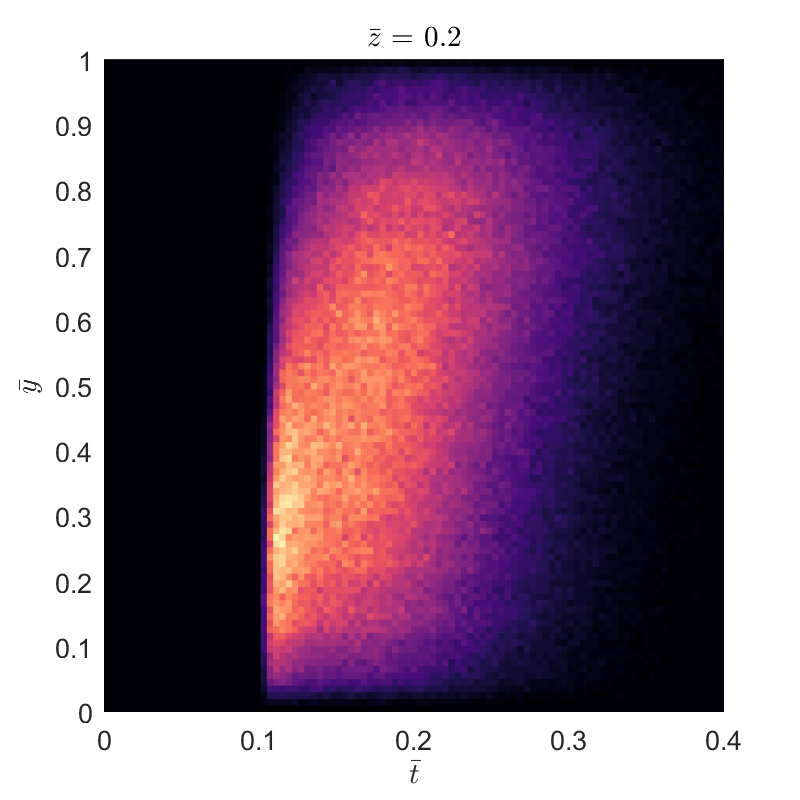}
        
        \vspace{-0.2cm}
        
		\caption{$\hat{z} = 0.2$}
		\label{subfig_histogram_cylinder_02}
	\end{subfigure}
    \begin{subfigure}{0.32\textwidth}
		\includegraphics[width=\textwidth]{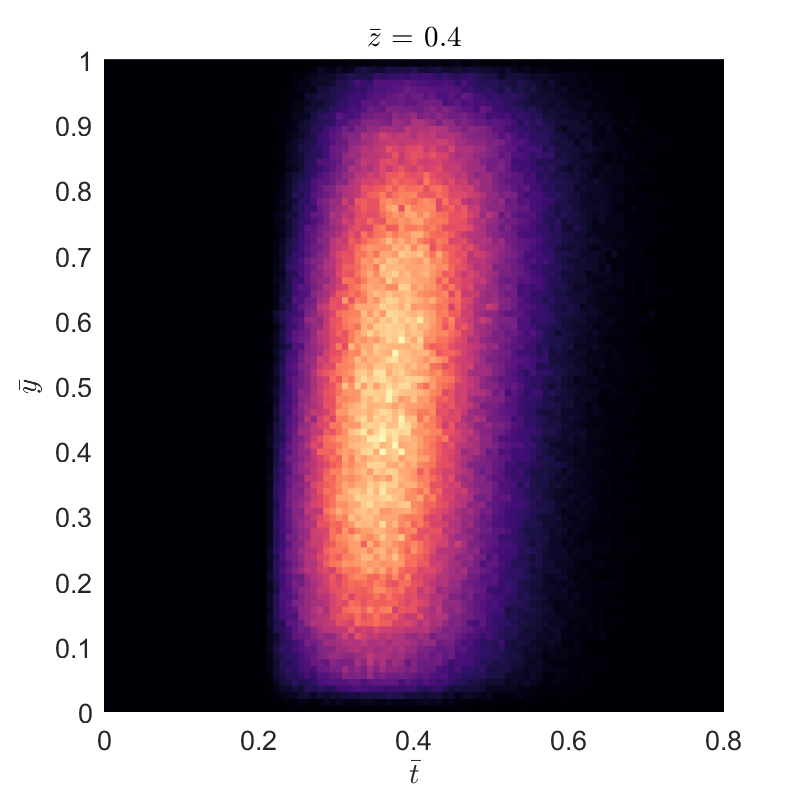}
        
        \vspace{-0.2cm}
        
		\caption{$\hat{z} = 0.4$}
		\label{subfig_histogram_cylinder_04}
	\end{subfigure}

        \vspace{-0.2cm}
        
	\caption{2D histograms for cylindrical channel geometries, constant flux source.}
	\label{fig_histogram_cylinder}
\end{figure*}

\begin{figure*}
	\centering
    \begin{subfigure}{0.45\textwidth}
		\includegraphics[width=\textwidth]{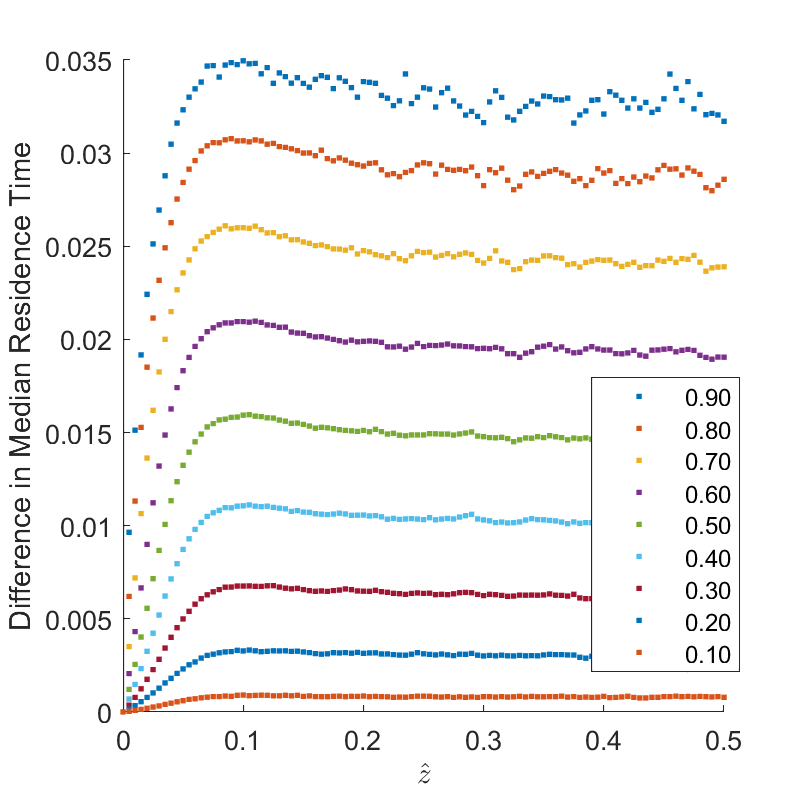}
        
        \vspace{-0.2cm}
        
		\caption{}
		\label{subfig_rtd_depths_cylinder}
	\end{subfigure}
	\begin{subfigure}{0.45\textwidth}
		\includegraphics[width=\textwidth]{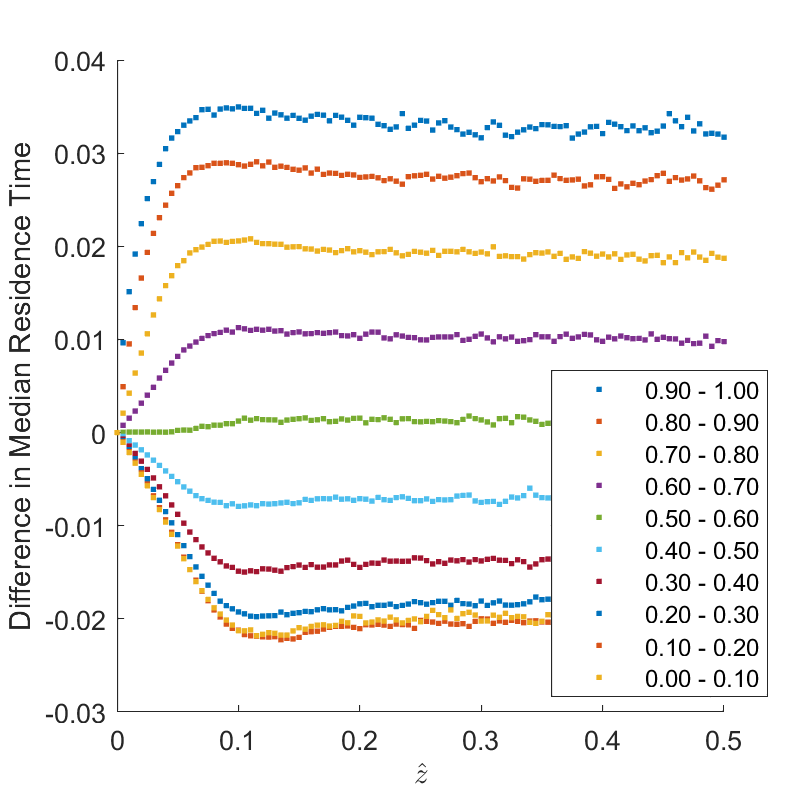}
        
        \vspace{-0.2cm}
        
		\caption{}
		\label{subfig_rtd_slices_cylinder}
	\end{subfigure}
        
        \vspace{-0.2cm}

	\caption{Median time difference for different penetration depths in a cylindrical channel. Constant flux source.}
	\label{fig_rtd_cylinder}
\end{figure*}

We next simulate the evolution of particles in a cylindrical channel. 2D histograms for different channel lengths are shown in Fig.\  \ref{fig_histogram_cylinder}. The older age of particles near the wall is particularly visible in this case, with the histogram's shape having a slight diagonal slant. The lower populations observed at lower $\eta$ in Fig.\  \ref{fig_histogram_cylinder} are an artifact of cylindrical coordinates, where areas of lower $\eta$ corresponds increasingly small circles.
We plot the effect for different depths, as illustrated in Fig.\  \ref{fig_rtd_cylinder}. The curves are qualitatively similar to the rectangular case.

\section{Practical Examples}\label{sec_practical_examples}

The results shown here have practical implications in real systems, and in some cases there is a risk of committing significant errors if they are not considered. As an example, one of the impetus behind this study was the characterization of experiments in the study of nanoparticle growth using flow dynamic light scattering (DLS) \cite{besseling2019new} systems.
%
A typical experiment may involve suspension of nanoparticles of radius of about 10 nm.
The diffusion coefficient of these nanoparticles can be approximated using the Stokes-Einstein relation,
\begin{equation}
    D = \frac{k_B T}{6 \pi \mu r},
\end{equation}
which yields $D \sim 2$$\times$$10^{-11}$ m$^2/$s at room temperature for a solvent with viscosity comparable to that of water.
%
%
We have shown in Section \ref{sec_results_cylinder} that the cutoff between the regions of compounding lag and the region of constant additional delay corresponds to roughly $\hat{z} \sim 0.1$ for a cylindrical channel. Plugging our typical values into equation \ref{eq_scalings}, this would correspond to a dimensional length of $z_\mathrm{transition} = 25$ m for a flow rate of $Q = \pi a^2 V = 100$ \textmu{l}/min. If we were to then interrogate particles in the $10 \%$ of the channel nearest to the wall, these would have a dimensionless median residence time difference of $\hat{t} \sim 0.033$ with the cross-sectional average (taken from Fig.\ \ref{subfig_rtd_depths_cylinder}), which for typical tubes with $r \sim 500$ \textmu{m} would here correspond to an enormous dimensional lag of $t \sim 12500$ s.
In this model experiment, if the tube length was shorter than the transition length of $25$ m (which it most likely would be), the difference in measured age of the particles would be a more or less linear function of the tubing length used, up to the maximum of $t \sim 12500$ s at $25$ m. For example, an experiment with 1 m of tubing would see a difference in particles' median age between the wall region and the entire cross-section of roughly $500$ s.
This is important to consider, as it means that the length of tubing used in an experiment can become a variable that significantly affects the measured results.

Further complicating the situation, if the measurement was taken less than $500$ s before the start of the experiment, then the difference in the median age of the particles between the region near the edge of the wall and the cross-sectional average would be a transient value that would depend both on channel length and on the moment the measurement was taken (see Appendix \ref{appendix_transient}).
In experiments that are probing kinetics of nanoparticles using finite penetration depth analytical techniques, Taylor dispersion is going to severely bias the results, such that the particles near the wall are not representative of the entire cross-section.
%



\section{Conclusion}

In conclusion, we have shown how residence time distributions of particles vary across the vertical distance in rectangular channel flows, as well as radial distance in cylindrical pipe flows. We have shown that particles near the wall accumulate "lag" up to a certain critical distance, after which diffusion counterbalances advection, and the particles have a constant delay when compared with the cross-sectional average. This effect becomes important in high Peclet number flows, with delays easily adding up to the order of minutes or even hours in typical scenarios.
The results presented are of particular importance in experiments probing kinetics (for example studying the growth mechanisms of nanoparticles) using measurement tools with finite penetration depths, for example, flow DLS systems.

\begin{acknowledgements}
This research is funded by the U.S. Food and Drug Administration under the FDA BAA-22-00123 program,
Award Number 75F40122C00200. Etienne Boulais acknowledges funding from the Fond de Recherche du Quebec, Nature et Technologie (FRQNT) Postdoctoral Fellowship Program, as well as National Science and Engineering Research Council of Canada (NSERC) Postdoctoral Fellowship Program
\end{acknowledgements}

\appendix

\section{Additional Results}\label{appendix_additional}

\begin{figure*}
	\centering
    \begin{subfigure}{0.45\textwidth}
		\includegraphics[width=\textwidth]{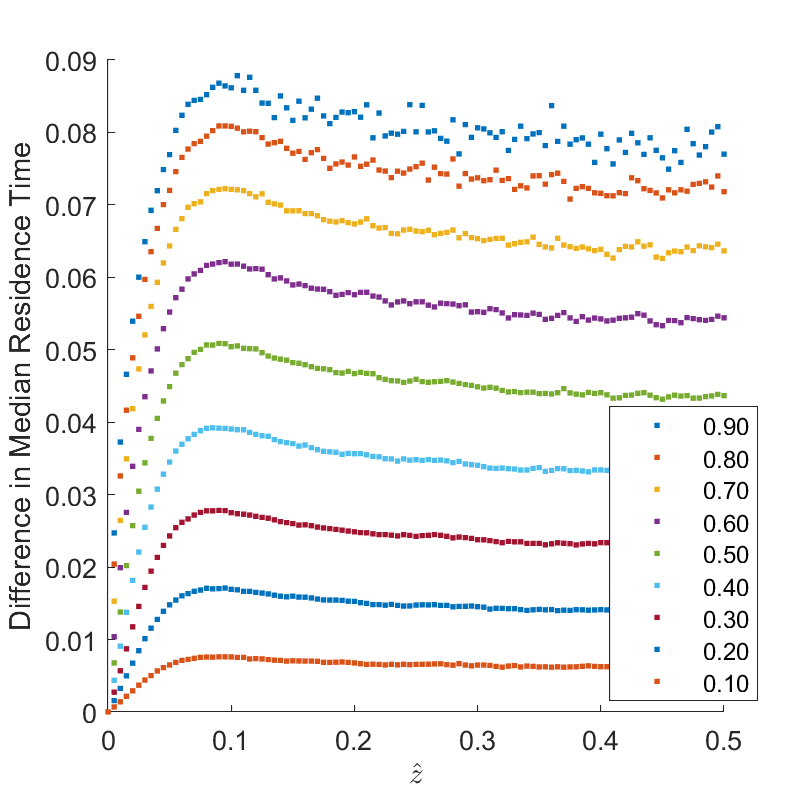}

        \vspace{-0.2cm}
        
		\caption{}
	\end{subfigure}
	\begin{subfigure}{0.45\textwidth}
		\includegraphics[width=\textwidth]{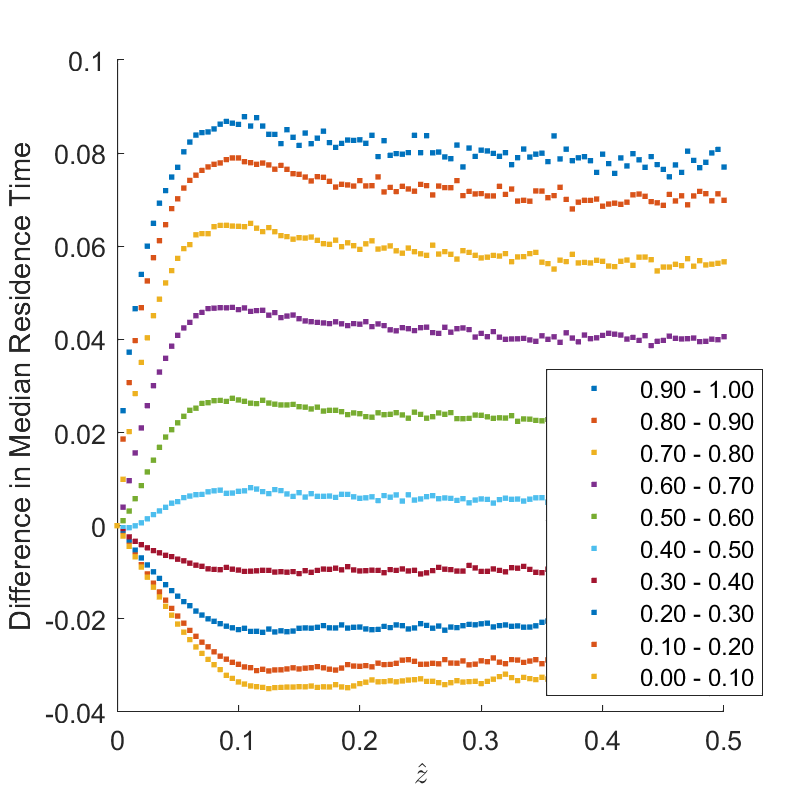}
        
        \vspace{-0.2cm}

		\caption{}
	\end{subfigure}

        \vspace{-0.2cm}
        
	\caption{Difference in dimensionless median time for different penetration depths $\abs{\hat{y}} < k$ (a), as well as for different slices $k_1 \leq \abs{\hat{y}} < k_2$ (b). 2D channel. Initial condition: uniform concentration plug.}
	\label{fig_rtd_depths_rect_uniform}
\end{figure*}

\begin{figure*}
	\centering
    \begin{subfigure}{0.45\textwidth}
		\includegraphics[width=\textwidth]{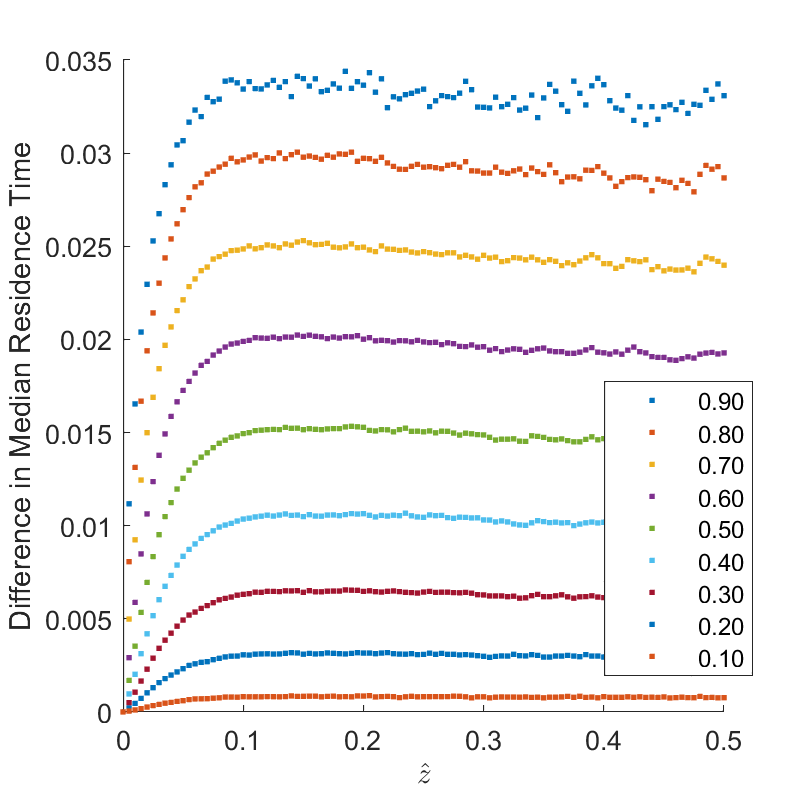}
        
        \vspace{-0.2cm}
        
		\caption{}
	\end{subfigure}
	\begin{subfigure}{0.45\textwidth}
		\includegraphics[width=\textwidth]{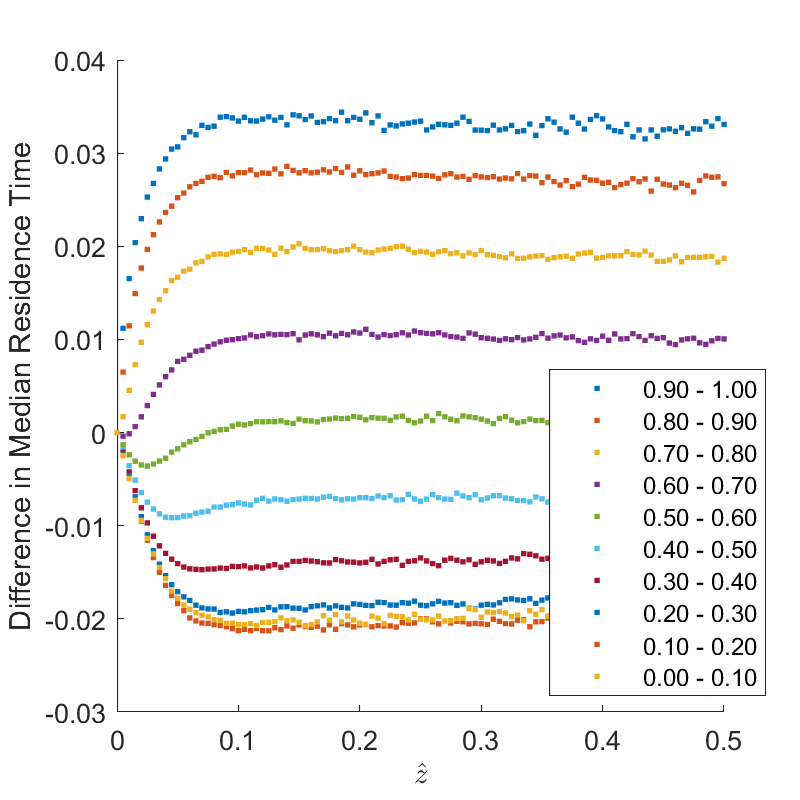}
        
        \vspace{-0.2cm}
        
		\caption{}
	\end{subfigure}

        \vspace{-0.2cm}
        
	\caption{Difference in dimensionless median time for different penetration depths $\abs{\hat{y}} < k$ (a), as well as for different slices $k_1 \leq \abs{\hat{y}} < k_2$ (b). Cylindrical channel. Initial condition: uniform concentration plug.}
	\label{fig_rtd_depths_cylinder_uniform}
\end{figure*}

The results in the main text were for an initial condition of constant flux through the cross-section at $\hat{z} = 0$. Here we present those same results for an infinitely thin plug of constant concentration at $\hat{z} = 0$. Results are shown for a 2D semi-infinite (Fig.\  \ref{fig_rtd_depths_rect_uniform}), and cylindrical (Fig.\  \ref{fig_rtd_depths_cylinder_uniform}) channels. 

\section{Transient Example}\label{appendix_transient}

\begin{figure}
	\centering
    \includegraphics[width=0.4\textwidth]{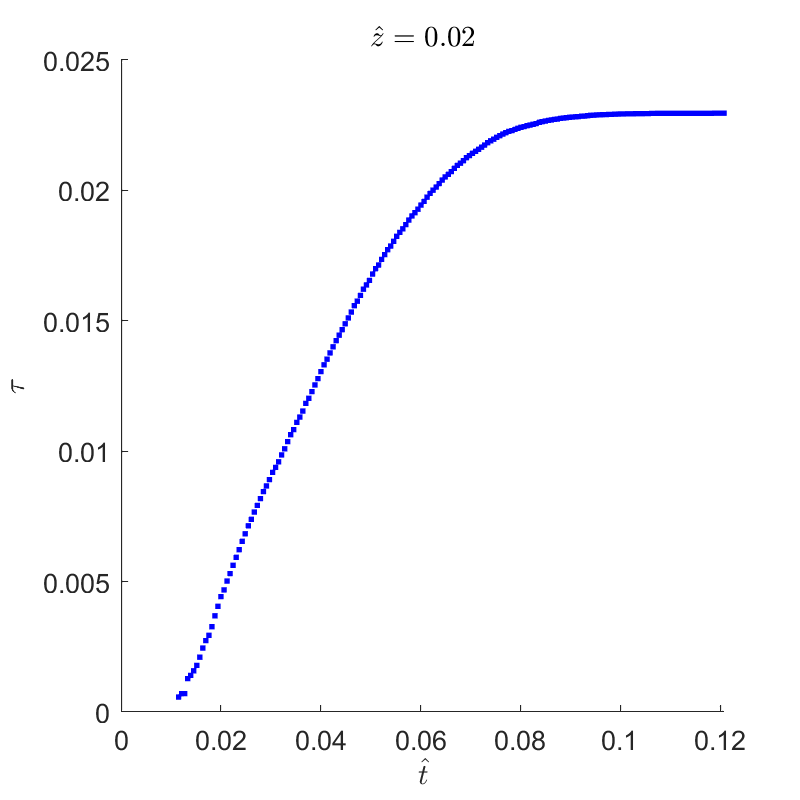}

        \vspace{-0.2cm}
        
	\caption{Difference in residence time between the region near the wall ($\eta > 0.9$) and the cross-sectional average in a cylindrical pipe at $\hat{z} = 0.02$ for short times (constant flux initial condition).}
	\label{fig_transient_delay}
\end{figure}

Section \ref{sec_practical_examples} provided examples of dimensional delays for particles near a wall in typical experiments. Such delays can easily add up to minutes or hours, once steady state has been achieved. If measurements are taken before that steady state is achieved, the delay will be shorter. Figure \ref{fig_transient_delay} shows an example of how that delay evolves for a set position within a cylindrical tube (here $\hat{z} = 0.02$). For short times, no particle is recorded (when convection has not brought in the first particles), then the difference in median times between the center and edge of the channel compounds until the whole tail has passed. In cases where the delays add up to hours, this means that the difference between the particles near the wall and the cross-sectional average might not just be a function of how far along the tube the measurement apparatus is, but also of how long has elapsed since the experiment started. Considering these effects is important when probing particle kinetics that are of a comparable time scale as the delays induced.

\bibliographystyle{unsrt}
\bibliography{taylor_dls}

\end{document}